\documentclass[twocolumn, prl]{revtex4-1}
\usepackage{graphicx}
\usepackage{epsfig}
\usepackage{color}
\usepackage{tabu}
\usepackage{braket}
\usepackage[colorlinks,linkcolor=blue]{hyperref}
\usepackage{amsmath}

\newcolumntype{C}{>{$}c<{$}} 
\begin{document}
\title{Manipulating anyons in quantum Hall droplets of light using dissipations 
}
\author{Yangqian Yan, Qi Zhou}
\affiliation{Department of Physics and Astronomy, Purdue University, West Lafayette, IN, 47907}
\date{\today}
\begin{abstract}

Whereas anyons are the building blocks in topological quantum computation, it remains challenging to create and control each anyon individually. Here, we point out that dissipative dynamics in cavities deterministically deliver droplets of light in desired fractional quantum Hall states. In these quantum Hall droplets, both the number and locations of anyons are precisely controllable without requiring extra potentials to imprint and localize such quasiparticles.  Using the density profile of light, the anyonic statistics is readily accessible. Moreover, entangling a quantum spin valve  with these quantum Hall droplets establishes a direct readout of the braiding statistics.  Our work unfolds a promising route for quantum optics to solve challenging problems in quantum Hall physics. 

\end{abstract}

\maketitle
It has been a long-lasting goal of physicists to detect anyonic statistics,
i.e., 
exchanging two quasi-particles leads to a phase that is neither 0 nor
$\pi$~\cite{Paredes2001,Kitaev2003,Nayak2008,Dutta2018}. Recent
experiments have reported direct observations of anyonic statistics in two-dimensional electron gases using either Fabry-Perot interferometers or collisions between anyons at a beamsplitter~\cite{Bartolomei2020,Nakamura2020}.  In spite of these exciting developments, to fully utilize anyons in quantum computation, it is required to have the capability to manipulate a single anyon in experiments. Such a task remains challenging in two-dimensional electron gases since it is difficult to precisely control the number of anyons in the sample. For instance, 
neither the number of anyons confined by the gates nor that in the edge currents is unambiguous in typical solid materials. It is thus desired to explore other platforms, in which anyons can be deterministically created and controlled.

Recent progress in quantum optics has provided physicists with
an unprecedented opportunity to explore quantum Hall
physics using highly controllable photon-atom
interactions~\cite{Jia2018,Ozawa2019,Clark2020}.
Synthetic Landau levels have been realized in twisted cavities to
access quantum Hall physics in both a flat plane or a cone~\cite{Lu2014,Schine2016,Jia2018a,Ozawa2019}.
In the latter case, only a fraction of the lowest Landau levels is occupied, for
instance, $z_j^{3m}$, where $z_j=(x_j+i y_j)$ is the complex coordinate of the
$j$th particle and $m$ is an integer. 
Strong
interactions between photons have been introduced by hybridizing photons with
atoms~\cite{Lukin2003,Birnbaum2005,Fushman2008,Peyronel2012,Chang2014,Firstenberg2016,Hartmann2016}.
Furthermore, two photons uploaded to a twisted cavity have been prompted to the
quantum Hall regime~\cite{Clark2020}. In parallel, 
it has been found that the internal state of a single atom in a cavity 
has been entangled with photons, enabling 
a new means 
of quantum spin-valve to control
hybridized photon-atom
systems~\cite{Reiserer2013}. 
Based on these currently available techniques, here, 
we point out a scheme to 
deterministically deliver desired 
quantum Hall droplets of light, in which both the number and locations of anyons are controllable.

\begin{figure} 
  \includegraphics[angle=0,width=0.5\textwidth]{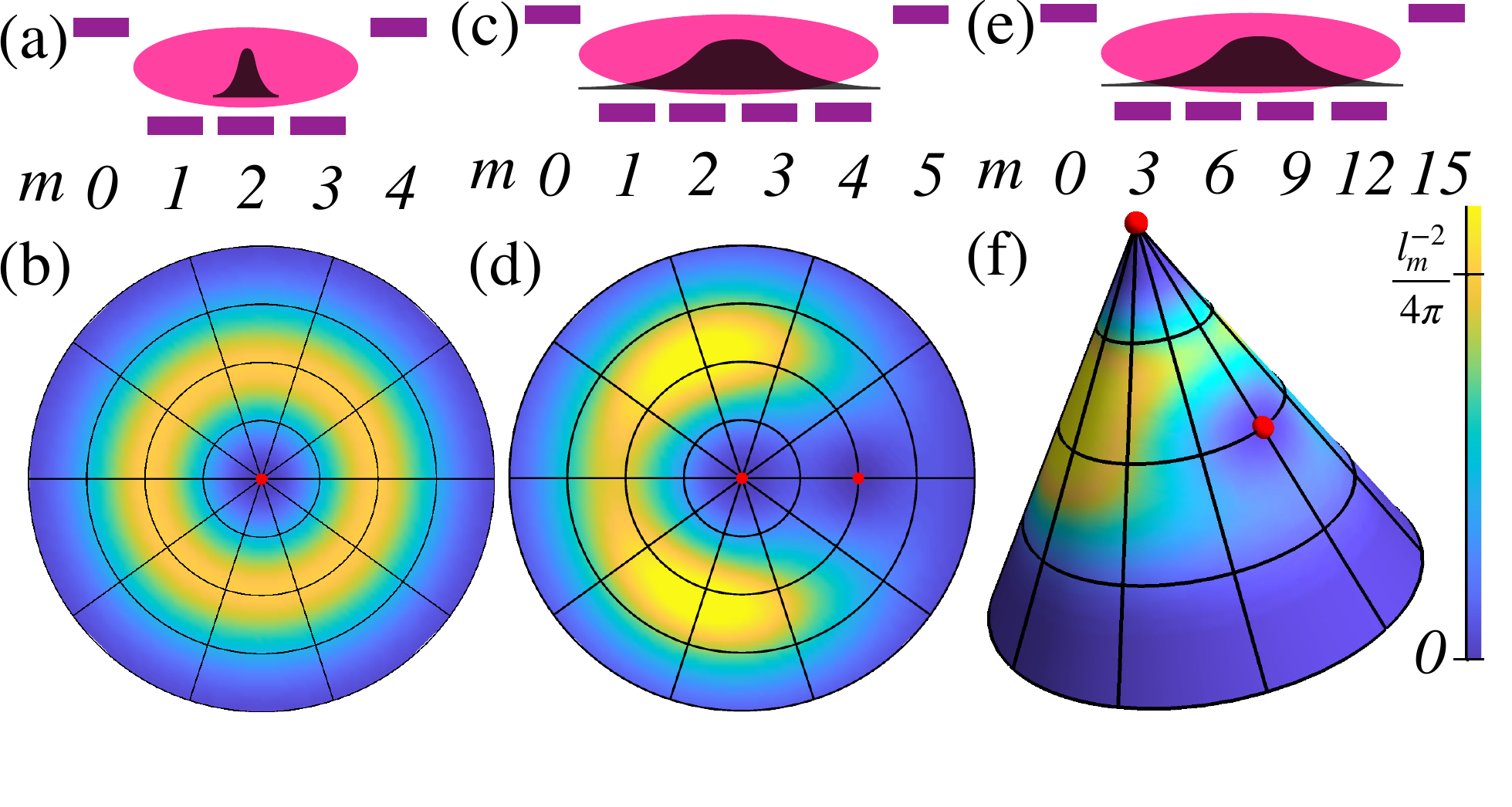}
  \caption{
    (a-b) An initial state of photons occupying a single LLL dissipates to a quantum Hall droplet with an anyon fixed at the origin. 
    An appropriate linear
    combination of multiple LLLs allows the initial state to dissipate to a quantum Hall droplet with two anyons in a flat plane (c-d) and a cone (e-f).
    The red dots mark the locations of anyons. Density distributions of the droplets are highlighted by colors. The allowed LLLs are constraint between $m_{min}$ and $m_{max}$. 
    } \label{fig1} \end{figure}

We considered a system identical to that in the Chicago experiment, where 
$N$ atoms are uploaded to a twisted cavity.  In this experiment, 
these photons are prepared at an initial state $\sim a^{\dagger N}_m|0\rangle$, where 
$a^\dagger_m$ is the creation operator of a photon at the lowest Landau level (LLL) with an angular momentum $m$.  Whereas a dissipative dynamics turns this trivial initial state into a one in the quantum Hall regime, only one anyon has been created and the position of the anyon is fixed at the origin. 
This scheme applies to both a cone and a flat plane as shown in Fig.~\ref{fig1}(a-b).  To levitate constraints 
to the number and location of anyons in this scheme, we consider an initial state  $\sim(\sum_m c_m
a^\dagger_m)^N|0\rangle$, i.e.,  
each photon is prepared at a superposition of multiple angular momenta. This can
be easily realized using a pumping laser with multiple angular momentum modes~\cite{[{For example: }] Fontaine2019}. 
Such a simple generalization leads to conceptual differences in the resultant quantum Hall droplets. For instance, two anyons could emerge in the droplet, 
one placed at the origin
of a flat plane [Fig.~\ref{fig1}(c-d)] or the apex of a cone [Fig.~\ref{fig1}(e-f)] and the other separated by a certain
distance. A unique advantage of this scheme is that external potentials are not required to localize anyons.  In fact, the position of the second anyon is automatically determined by $\{c_m\}$, i.e., the initial condition of the dissipative dynamics. As such, by changing $\{c_m\}$, the position of the second anyon is highly tunable. 

We define 
$\ket{vac}$ as the state with all atoms occupying the $s$ orbital and  no photon in the cavity. 
The time evolution of
the density matrix, $\rho$, of this open system is determined by 
a master equation~\cite{Cohen-Tannoudji1992, Clark2019},
$\dot\rho=\frac{i}{\hbar}[\rho,H_0+H_{pump}+U]+\mathcal{D}[\rho]$, where
\begin{align}
 H_0 &=
 \sum_{m} (g p_m^\dagger a_m
  +\Omega r_m^\dagger p_m +h.c.), 
\label{ham}\\
  H_{pump}&= \sum_m\Omega_{m}^{pump} \left( a^\dagger_m + a_m \right).\\ 
  \mathcal{D}[\rho]&= 
  \sum_m  \gamma
  \left(  p_m \rho p_m^{\dagger} -\frac{1}{2}\{\rho, p_m^{\dagger} p_m\} \right)
  \label{master}
\end{align}
Replacing $m$ by $3m$, the master equation applies to a cone. The first term in
Eq.~(\ref{ham}) describes a resonant coupling of the cavity mode with the
atoms. To simplify
notations, we have used $p^\dagger_m$ ($r^\dagger_m$) and $p_m$ ($r_m$) to denote the atomic transitions between the $s$ orbital and the short-lived $p$ orbital (long-lived Rydberg orbital). We have also considered a constant coupling, $g$. Our results can be
directly generalized to $m$-dependent couplings.  The second term in
Eq.~(\ref{ham}) is induced by an external laser that couples the $p$ state to the Rydberg state with a coupling strength $\Omega$.  
The subscript of the operators, $m$, denotes collective excitations of atoms. For instance,
${a}_m^\dagger=\int_z dz  \varphi_m(z) {a}_z^\dagger$, 
where $z=(x+i y)/\ell_m$ is the real space coordinate,
$\varphi_m(z)=\frac{1}{\sqrt{\pi m!} \ell_m} z^{m} e^{-|z|^{2} / 2}$ is the wavefunction with an angular momentum $m$ in the LLL, and ${a}^\dagger_z$ is the operator for photons  at position $z$. 
The magnetic length $\ell_m$ has been set to 1.  
${p}_m^\dagger$ and ${r}_m^\dagger$ can be expressed in terms of operators in the real space in a similar manner. 

Because of the diluteness of the atoms inside the cavity, the interaction
between two Rydberg atoms is well modeled by a contact
interaction $U$~\cite{Popp2004} and the interactions between two atoms at the $s$ and $p$ orbitals are negligible,
\begin{align}
U=U_0\sum_{\{m_i\}}U_{m_1,m_2}^{m_3,m_4} r_{m_1}^\dagger r_{m_2}^\dagger r_{m_3}
r_{m_4},\\
U_{m_{1},m_{2}}^{m_{3}, m_{4}}=\frac{\left(m_{1}+m_{2}\right)!}{2\pi2^{m_{1}+m_{2}} \sqrt{m_{1}! m_{2}! m_{3}! m_{4}!}}.
\end{align}
In Eq.~(\ref{master}), $\gamma$ describes the decay of atoms in the short-lived $p$ orbital, and $\Omega_{pump}$ denotes the strength of an additional field that continuously pump photons to the cavity. 
Again, $\gamma$ has been treated as a constant for simplicity.  
We first consider a vanishing pump, 
$\Omega_m=0$.  
The initial  state, $|\Psi_{\text I}
\rangle=(\sum_m c_m
a^\dagger_m)^N|vac\rangle/\sqrt{N!}$, 
can be rewritten in the coordinate space as 
\begin{equation}
|\Psi_{\text I}\rangle=\mathcal{N}_{\text I}^{-\frac{1}{2}}\int \prod_idz_i \Psi_{\text I}(\{z_i\}) a^\dagger_{z_i}|vac\rangle,
\label{eqinitialstate}
\end{equation}
where $\Psi_{\text I}(\{z_i\})=\prod_{i=1}^N(\sum_m{c}_m \varphi_m(z_i))$, $\{z_i\}$ is a short-hand notation for $\{z_1,z_2,...z_N\}$, and
$\mathcal{N}_{\text I}$ is the normalization factor. For $N=1$, it is easy to see that all bright
states vanish at long times, and the initial state decays to a
one-body zero-energy dark polariton state,  
\begin{align}
&\rho_{\text I}=|\Psi_{\text I}\rangle\langle \Psi_{\text I}| \stackrel{t\gg
\gamma^{-1}}{\xrightarrow{\hspace*{0.6cm}} }\rho^o_{\text F}= |\Psi_{\text F}^o\rangle\langle\Psi_{\text F}^o|,\\
& |\Psi_{\text F}^o\rangle=\lambda\int dz\Big(\sum_m
c_m\varphi_m\Big)d^\dagger_{z} |vac\rangle,
\end{align}
where $\lambda=\frac{\Omega}{\sqrt{\Omega^2+g^2}}$ and $d^\dagger_z= \lambda
a_z^\dag -\frac{g \lambda}{\Omega} r_z^\dag $ is the creation operator for a
dark state at $z$ (Supplemental Materials). Since the $p$ orbital is not involved, $|\Psi_{\text F}^o\rangle$ as a superposition of eigenstates of $H_0$ survives the dissipative dynamics. 

When $N>1$, a $N$-polariton zero-energy dark states, $|D_l\rangle$,
is labeled by $l$,  the angular momentum that is conserved in the dissipative dynamics. 
$|D_l\rangle$ 
satisfies two criteria. First, it must be constructed by only 
one-body dark polariton states. Otherwise, the $p$ orbital will be present and causes decay. Second,
its spatial wavefunction
must
vanish when two dark polaritons sit on top of each other. Otherwise, the finite interaction energy leads to an off-resonant condition such that this state cannot be excited. As such, in the long time limit, the dissipative dynamics in the cavity amounts to projecting the initial state to a superposition of $|D_l\rangle$ with zero-energies,
\begin{align}
&|\Psi_{\text F}\rangle=\sum_l \tilde{u}_l |D_l\rangle,\\
&|D_l\rangle=\int \prod_{i=1}^N dz_i \phi_l(\{z_i\}) \prod_id_{z_i}^\dagger|vac\rangle, \label{md}
\end{align}
where $\tilde{u}_l=\langle D_l|\Psi_{\text I}\rangle$~\cite{Julia-Diaz2012a},
and $\phi_l(\{z_i\})=0$ when $z_j=z_k$ for any $j\neq k$ (Supplemental Materials). The above discussions show that, if an appropriate $|\Psi_{\text F}\rangle$ is chosen, the corresponding $u_l$ could deliver a target state of interest.

As an example, we  show 
how to obtain a $N$-polariton state in a flat plane with two anyons, i.e., two quasi-holes, 
\begin{align}
&|\Phi\rangle=\int \prod_idz_i\Phi(\{z_i\})\prod_i d^\dagger_{z_i}|vac\rangle,\\
&\Phi(\{z_i\})=\mathcal{N}^{-\frac{1}{2}}\prod_{i=1}^Nz_i(z_i-\eta)\prod_{i<j}(z_i-z_j)^2\prod_i e^{-\frac{|z_i|^2}{2}},\nonumber
\end{align}
where

the first anyon is located at $z=0$ and the other at $z=\eta$. 
All results can be immediately generalized to a cone by replacing $z_i$ by $z_i^3$. 

Expanding $\Phi(\{z_i\})$ in terms of $\eta$, we see that it is a superposition of multiple zero-energy states, 
\begin{equation}
  \Phi(\{z_i\})=\mathcal{N}^{-\frac{1}{2}}\Big(\sum_{l=N^2}^{N^2+N}(-\eta)^{N^2+N-l}\phi_l\Big),\label{quasihole}
\end{equation} 
where
\begin{align}
  &\phi_{N^2}(\{z_i\})=\prod_{i=1}^Nz_i\prod_{i<j}(z_i-z_j)^2\prod_ie^{-|z_i|^2/2},\\
&\phi_{l}(\{z_i\})=\left(\sum_{i_1,i_2,...,i_{N-(N^2-l)}} \prod_{k=1}^{N-(N^2-l)}z_{i_k}\right)\phi_{N^2}(\{z_i\}.\label{ze})
\end{align}
Apparently, each $\phi_l(\{z_i\})$ vanishes whenever two
particles have identical coordinates and thus have a zero interaction energy. It carries a unique total angular momentum $l$,
and $\Phi$ is a superposition of multiple states with different angular momenta,
$l=N^2, N^2+1, ..., N^2+N $, when $\eta\neq 0$. Interestingly, each
$\phi_{l}(\{z_i\})$ is precisely a Jack polynomial, a powerful tool that has
been extensively studied in quantum Hall physics (Supplemental Materials).

To access $|\Phi\rangle$ at long times in the dissipative dynamics, we required that the allowed single-particle states in the LLL are $m=1,2,\dots,N$.   As such, 
the initial state is written as
\begin{equation}
  \Psi_{\text I}(\{z_i\})=\mathcal{N}_{\text
  I}^{-\frac{1}{2}}\prod_{i=1}^{N}\left(z_i^{(N+1)}+\sum_{m=1}^{N} c_m
  {z_i^{m}}\right)\prod_i e^{-|z_i|^2/2},\label{psiI}
\end{equation}
which depends on $N$ parameters. 
Apparently,  $\Psi_{\text I}$ is also a
superposition of multiple states with angular momenta $N$, ..., $N^2+N$,
\begin{equation}
\Psi_{\text I}=\sum_{l=N}^{N^2+N}  \psi_l (\{z_i\}).\label{PisiI}
\end{equation}

When $l<N^2$, there does not exist a
state in LLL with vanishing contact interaction energy as long as
single-particle $m=0$ states are
excluded. Thus, any
$\psi_{l<N^2} (\{z_i\})$ can only be turned into bright states and eventually be
dissipated away.  When $N^2\le l\le N^2+N$,  
$\phi_l(\{z_i\})$ is only zero energy eigenstate for each $l$,
provided that the available single-particle states are constrained between $m_{\text{min}}=1$ and $m_{\text{max}}=N+1$. The cutoffs, $m_{\text{min}}=1$ and $m_{\text{max}}=N+1$, can be provided by either the finite size of the mirror
or purposely limiting available angular momentum channels~\cite{Clark2019}, as shown in Fig.~\ref{fig1}. Therefore, $\psi_l (\{z_i\})$, which has a fixed angular momentum $l$, must decay to $\phi_l(\{z_i\})$ in long times.

We denote the overlap between $\psi_l (\{z_i\})$ and $\phi_l$ as $\mu_l$,
\begin{equation}
  u_{l}=\frac{\int \prod_i dz_i \phi^*_l(\{z_i\})\psi_l (\{z_i\})}{
  \int \prod_i dz_i \phi^*_l(\{z_i\})\phi_l (\{z_i\})}
\end{equation}
To ensure that $\Psi_{\text I}(\{z_i\})$ in Eq.~(\ref{PisiI}) eventually dissipates to $\Phi(\{z_i\})$
as shown in Eq.~(\ref{quasihole}), it is required that 
\begin{align}
u_{l}(c_1,c_2,...,c_N)&=-\eta u_{l+1}(c_1,c_2,...,c_N), \\
 l&=N^2, ..., N^2+N-1. \label{eta}
\end{align}
Since there are in together $N$ such equations, we could then uniquely
determine the $N$ unknown parameters in Eq.~(\ref{psiI}).  For instance, if
$N=2$, we find that $c_1=-4\eta^2/25$, $c_2=-9\eta/10$.  A few other examples of
the explicit expressions of $c_m$ are given in the supplemental materials.  In
fact, $\psi_l (\{z_i\})$ in the initial state can also be written in terms of
Jack polynomials so as to simplify numerics (Supplemental Materials). When
Eq.~(\ref{eta}) is satisfied, we conclude that 
\begin{align}
|\Psi_{\text I}\rangle \stackrel{t\gg
\gamma^{-1}}{\xrightarrow{\hspace*{0.6cm}} }
\lambda^N\mathcal{N}_{\text I}^{-\frac{1}{2}}\mathcal{N}^{\frac{1}{2}}|\Phi\rangle,
\end{align}
i.e., the dissipative dynamics turns initial state into desired quantum Hall states.
Here the $\lambda^N$ factor originates from the overlap between the single-particle photonic state and the
dark-polariton state.

The same scheme could be applied to a quasi-hole state of $N$ polaritons on a
cone by tuning $\delta_m$, or the creation of a single quasi-hole state at
location $\eta$ (Supplemental Materials).

\begin{figure} 
  \includegraphics[angle=0,width=0.5\textwidth]{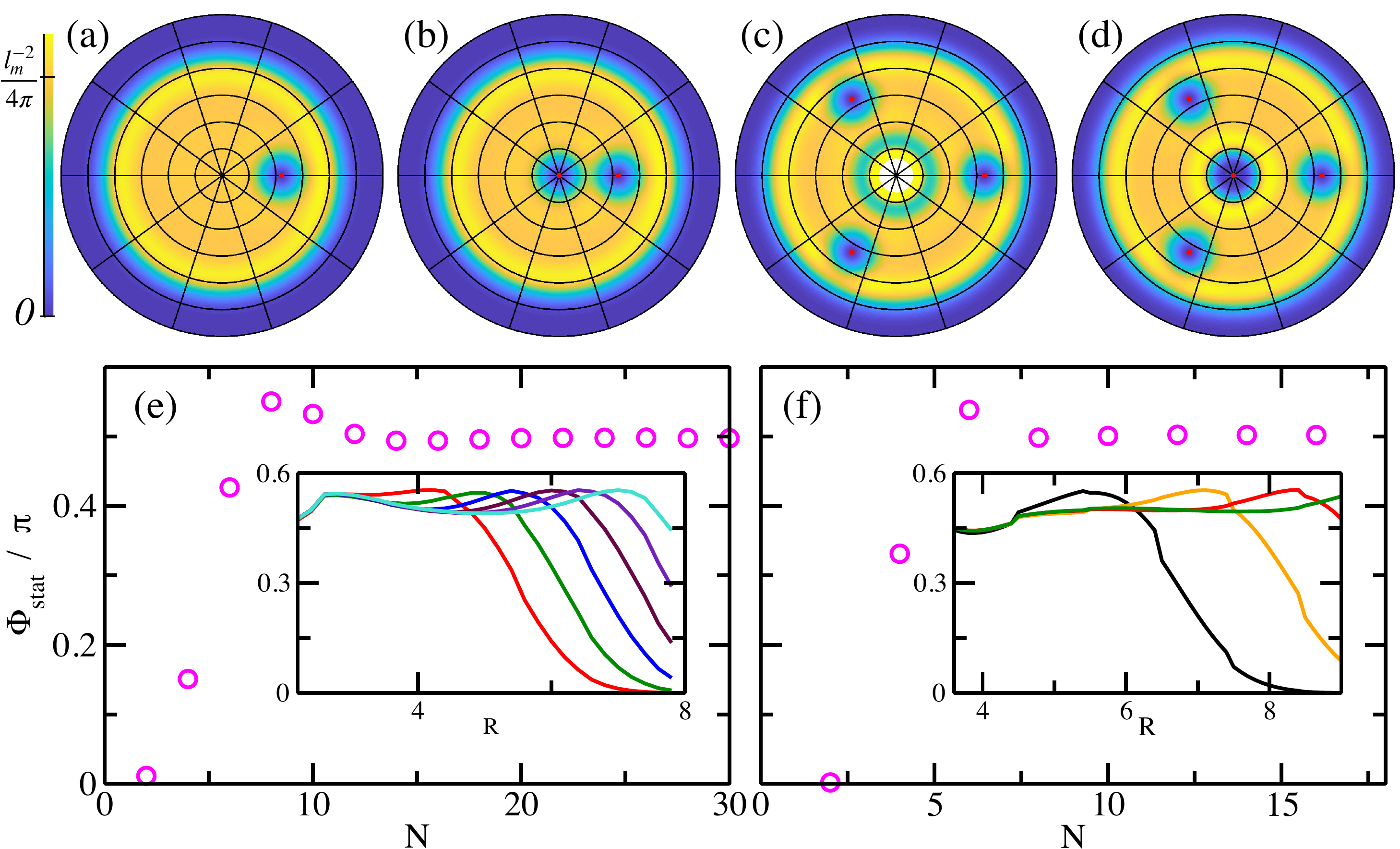}
  \caption{
    Radial density $\rho(r)$ for one-quasihole (ac) state and two-quasihole (bd) state.
    Panel (ab) are for the flat space and (cd) are for a cone with $\delta m=3$.
    The radial spacing is $2l_m$.
    The density is normalized to the number of particles $N=20$ ($N=8$) for panel
    (ab) [(cd)].
    Panel (e) [(f)] show the statistical phase factor as a function of $N$ in a
    flat space (cone) while fixing $r=4.4$ ($r=6.6$).
    Insets show the statistical phase factor as a function of $r$.
    Black, orange, red, green, blue, brown, purple, cyan lines are for
    $N=4,6, 8,10,12,14,16,18$.
    The red dots mark the locations of anyons.
      Due to the curvature singularity at the cone tip, the density at the origin of panel (c)  reaches
      $5.6l_m^{-2}/(4\pi)$, which exceeds the color scale. Thus, this region is
excluded from the density plot.
  } \label{fig2} \end{figure}

The dissipative dynamics at long times as discussed above can be verified by
numerically solving the master equation, from which the density matrix at any
given time is obtained. Since the numerics is heavy for $N>2$, we focus on $N=2$ when solving the master equation. 
We do observe that starting from a $2$-body state,
$( a_1^\dagger)^2|vac\rangle$, the final $2$-body state that survives the
dissipative dynamics is indeed the quasihole state at the origin (Supplemental Materials). Nevertheless,
we would like to point out that the full solution of the master equation
includes a mixture of density matrices with different total particle numbers. In
addition to the $N$-body sector, other sectors with particle numbers ranging
from $0$ to $N$ also exist.  The same analysis can be performed for each sector,
which hosts certain quantum Hall states with less than $N$
particles.
Alternatively, a simpler solution is to turn on $\Omega^{pump}$. 
The steady state is a superposition of states with the number of photons centered around
$N$~\cite{[{It is known
that the interplay between the pumping field and the dissipation could establish
a quasi-equilibrium in a cavity such that the number of photons or polaritions
is concentrated around a narrow peak at $N$ with variance $\sqrt{N}$. }]
Fox2006}.
Choosing the appropriate single momentum cutoffs, the $N+i$
photon sector with $i>0$ does not have zero energy state, under strong
interaction and large $p$-wave decay, they will decay thus $N$-photon event must
come from the $N$-body sector.

\begin{figure} 
  \includegraphics[angle=0,width=0.5\textwidth]{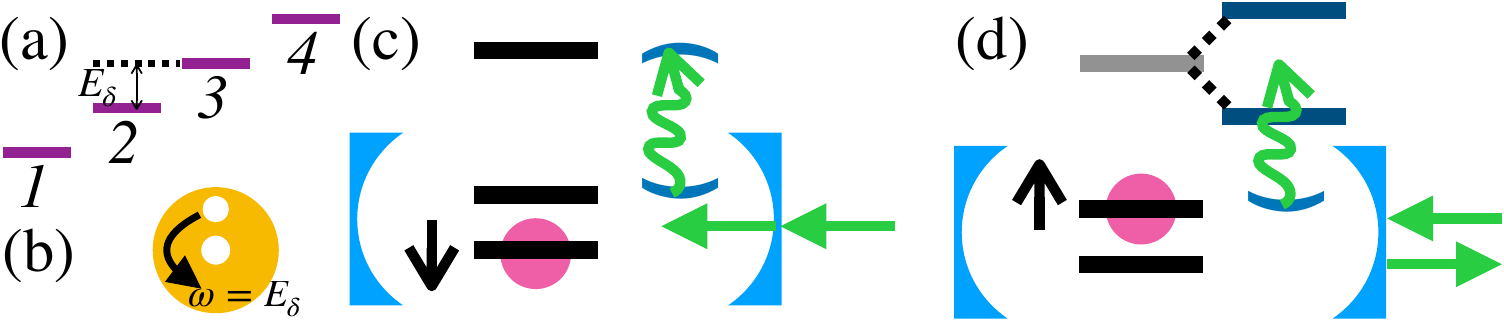}
  \caption{(a) Tilting the single-particle angular momentum levels results in
    the rotation of hole (b).
    (c-d) illustrates the coupling of a quantum spin valve with a many-body
    state. (c) While the spin is down, light can enter the cavity.
    (d) While the spin is up, light is reflected off due to off-resonance
    condition.
    As the number of particles increases, the statistical phase factor
    approaches $0.5\pi$. Inset shows the statistical phase factor as a function
    of $R$ for $N=20$.
   } \label{fig3} \end{figure}

We now turn to the measurement of anyonic statistics. The simplest way is to
measure the density profile of the quasi-hole state. It is known that moving a
single hole around a loop, the many-body wavefunction accumulates a phase
proportional to the total particle number enclosed by the
loop with a radius $R$~\cite{Arovas1984}. Adding an extra hole inside this loop, half of the
change of such geometric phase precisely provides us with the statistic phase of
the anyons,
\begin{align}
\phi_{\text{stat}}=\int_{0}^{R}\pi (\rho_{1,\eta}(r)-\rho_{2,\eta}(r)) d r . \label{phase},
\end{align}
where 
$\rho_{1,\eta}$ is the density of the state with one hole located at $z=\eta$
and $\rho_{2,\eta}$ is that of the state with an additional hole at the
origin~\footnote{ The density is defined as $\rho_{j,\eta}(r)=\int_0^{\infty} |\Psi_{j,\eta}|^2 \sum_i\delta(|z_i|-r)\prod_{i=1}^{N}dz_i$. The normalization gives $N$, i.e.,
$\int_0^{\infty} \rho_{i,\eta}=N$. The geometric phase difference is 
$\Delta\phi=\int_{0}^{R}2\pi (\rho_{1,\eta}(r)-\rho_{2,\eta}(r)) d r$,
Circling one hole around another by a loop is equivalent to exchange twice,
i.e., $\phi_{\text{stat}}=\Delta\phi/2$.
}. 
Thus, comparing the density profiles of these two quantum Hall states 
readily allows us to extract the abelian
statistic phase. In practice, however, a key question is the finite size effect
as Eq.~(\ref{phase}) is obtained in the thermodynamic limit where the size of
the hole is negligible. As shown in Fig.~\ref{fig2}, on both the flat space and
the cone, when $N=20$, the density at the plateau is readily close to $1/2$,
indicating that such particle number is readily a good approximation for the
results in the thermodynamic limit. The inset of Fig.~\ref{fig3}(c)
shows the
statistical phase factor over $\pi$ as a function of cutoff radius $R$. 
A plateau where anyonic statistics is robust is already clear. A more rigorous means is to trace the
dependence of the height of the plateau as a function of $N$, from which the
result at $N\rightarrow \infty$ can be extrapolated [Fig.~\ref{fig3}(c)]. 
Considering the cone leads to smaller required $N$~\cite{Wu2017} [Fig 3(d)].
For example, 
$\phi_{\text{stat}}$ on a cone shows a plateau at
$N=8$ and 10. For $N=4$ and $N=6$, we can already find a regime of $r$ where
$\phi_{\text{stat}}$ is close to 0.5.

A more direct means of measuring the anyonic statistics is to braid two anyons,
i.e., moving one around the other.  
The anyonic statistics encoded by the overlap between two many-body wavefunctions, one with anyons braided and the other without such a braiding, is then transferred to a quantum spin-valve.
Similar ideas of using a single spin to measure quantum coherence in many-body
systems have been studied for various purposes~\cite{Quan2006,Hanson2008,Wei2012,Reiserer2013,Vasilyev2020}. 
Here, measuring the spin coherence of the quantum spin-valve, the anyonic statistics can be extracted.
Whereas this scheme is very generic, here, we consider a quantum spin-valve in a cavity to concretize the discussions~\cite{Reiserer2013}. To this end, we consider the quasi-hole state
leaked out from the cavity where it is created. Then only the photonic part of
the state is relevant and the spatial part of the wavefunction remains
unchanged. Such quasi-hole state of photons is transferred to another cavity,
which includes a quantum spin-valve. This quantum spin-valve is made of a single
atom, whose hyperfine spin state controls whether the quantum Hall state of
photons could enter the cavity. When it is spin-up (down), photons can (cannot)
enter the cavity, similar to the experiment in reference~\cite{Reiserer2013}.
Thus, when the quantum spin-valve is prepared at a superposition of up and down,
two copies of the quasi-hole state of photons is created, 
$|\Phi\rangle=\frac{1}{\sqrt{2}}\left(|\uparrow\rangle|\Phi_\eta\rangle+\right|\downarrow\rangle|\Phi_\eta\rangle)$,
where the copy combined with spin-up(down) is inside (outside) the cavity.  

Inside the cavity, we now tilt LLL such that it is no longer flat and the single
particle energy becomes  $\sum_m m E_\delta a^\dagger_ma_m$, where
$E_\delta$ is a constant energy.  After tilting LLL for a certain time $t$,
each $\phi_l$ in Eq.~(\ref{ze}) acquires a dynamical phase, $e^{-i E_\delta l t}$.
This phase can then be absorbed by $\eta$ such that $\eta\rightarrow \eta'=\eta
e^{-i \omega t}$.  Thus, as time goes by, one hole encircle the other with an
angular frequency, $\omega=E_\delta$, without requiring  an external force to drag it,
as shown in Fig.~\ref{fig1}(d).  Meanwhile, the copy outside the cavity remains unchanged. After a certain time $t$, the wavefunction becomes
$|\Phi'\rangle=\frac{1}{\sqrt{2}}\left(|\uparrow\rangle|\Phi_{\eta'}\rangle+\right|\downarrow\rangle|\Phi_\eta\rangle)$
At this time, the coherence of the quantum spin-valve is given by
$\langle \downarrow |\uparrow\rangle=\frac{1}{2}\mathcal{F}\equiv\frac{1}{2}
\langle \Phi_{\eta}|\Phi_{\eta'}\rangle$.
In other words, the overlap between two quasi-hole states, $\mathcal{F}$, is
encoded in the quantum spin-valve. In particular, when $\eta$ and $\eta'$ are
close to each other, i.e., $t$ is small, the Berry connection can be obtained
from $\mathcal{F}$. In this particular braiding scheme, only the amplitude of
$\eta$ changes, and $\vec{A}=A\hat{\theta}$, where $\theta=\arg\eta$,
$\hat{\theta}$ is the unit vector in the angular direction, and $A=i\langle
\Phi_{\eta}|\nabla_\theta |\Phi_{\eta}\rangle/|\eta|$. It is straightforward to see that ${A}=i(1-\mathcal{F})/(\arg\eta-\arg\eta')$. Once $\vec{A}$ is obtained for each $\theta$, a loop integral of $\vec{A}$ then delivers the total geometric phase accumulated by braiding the anyons. 

To summarize, we propose to use dissipation to transform a 
trivial product state with mixed angular momentums to an entangled few-body
 Laughlin states with quasihole excitations. Tuning the initial mixing of the
angular momentum states, up to two quasiholes could be generated. The anyonic
statics could then be verified indirectly by measuring the density distribution
or directly by braiding. We hope the few-body version of the non-abelian anyonic state could be
engineered in the future.

\begin{acknowledgments}
This work is supported  by the Air Force Office of Scientific Research under
award number FA9550-20-1-0221 and a
seed  grant from PQSEI.
\end{acknowledgments}

\bibliographystyle{apstest}
\bibliography{Laughlin,note}

\clearpage
\pagebreak
\newpage
\widetext
\begin{center}
\textbf{\large 
Supplemental Material of ``Manipulating anyons in quantum Hall droplets of light using dissipations''
}
\end{center}

\renewcommand{\theequation}{S\arabic{equation}}
\renewcommand{\thefigure}{S\arabic{figure}}
\renewcommand{\thetable}{S\arabic{table}}
\renewcommand{\theHequation}{Supplement.\theequation}
\renewcommand{\theHtable}{Supplement.\thetable}
\renewcommand{\theHfigure}{Supplement.\thefigure}
\setcounter{table}{0}
\setcounter{figure}{0}
\setcounter{equation}{0}

\onecolumngrid

In this supplemental material, we present results of the one-body dark state, Jack polynomials for both one and two quasi-hole states, mapping initial state to monimial basis, and solutions to the master equation.

\section{The one-body dark state}

The spatial wave function in first quantization form reads
$\Psi_{\text I}(\{z_i\})=\prod_{i=1}^N(\sum_m{c}_m \varphi_m(z_i))$.
Though 
Eq.~(3) from the main text
can be solved numerically, it is
useful to analytically study what state survive the dissipative dynamics at long
times.
When $N=1$,  we see from 
Eq.~(1) from the main text
that the eigenstate of $H_0$ can be written as $ |D_m\rangle=d^\dagger_m\ket{vac}$, $|B_{m,\pm} \rangle=b^\dagger_{m,\pm}\ket{vac}$, and 
\begin{align}
& d^\dagger_m= \lambda a_m^\dag -\frac{g \lambda}{\Omega} r_m^\dag , \nonumber \\
&b^\dagger_{m,\pm}=\frac{g \lambda}{\sqrt{2}\Omega}a_m^\dag +\frac{\lambda}{\sqrt{2}} r_m^\dag \pm \frac{1}{\sqrt{2}}p_m^\dag.
\end{align}
where $\lambda=\frac{\Omega}{\sqrt{\Omega^2+g^2}}$. 
Apparently, the dark state, $|D_m\rangle$, of a polariton is immune to the dissipation as it does not contain the
$p$ orbital. 
The two bright states of polaritons, $|B_{m,\pm}\rangle$, eventually decay and
will not contribute to the final density matrix when $t\gg 1/\gamma$. If we use $|\Psi^{1}_{\text I}\rangle$ to denote the initial state of a single particle, 
all bright states vanish at long times, and 
\begin{equation}
|\Psi^{o}_{\text I}\rangle \stackrel{t\gg \gamma^{-1}}{\xrightarrow{\hspace*{0.6cm}} } \lambda\int dz\Big(\sum_m c_m\varphi_m(z)\Big) d^\dagger_{z}|vac\rangle,
\end{equation}
where 
$d^\dagger_m=\int dz \varphi_m(z)d^\dagger_z$, and $d^\dagger_z=\sum_m \varphi_m^*(z)d^\dagger_m$. 

\section{Jack polynomials and quantum Hall states}

The wave function of the quasihole state carrying a single quasihole is written as  
\begin{align}
&\Phi_{1,\eta}(\{z_i\})=\mathcal{N}^{-\frac{1}{2}}_{1}\prod_{i=1}^N(z_i-\eta)\prod_{i<j}(z_i-z_j)^2\prod_i e^{-\frac{|z_i|^2}{2}},
\label{oh}
\end{align}
where

$\eta$ denotes the location of quasihole.  When $\eta$ is finite, this state is a superposition of multiple angular momentum states. 
Expanding Eq.~(\ref{oh}) in power series of $\eta$, we obtain
\begin{equation}
  \Phi_{1,\eta}(\{z_i\})=\mathcal{N}_1^{-\frac{1}{2}}\Big(\sum_{l=N^2-N}^{N^2}(-\eta)^{N^2-l}\phi_l(\{z_i\})\Big)\prod_ie^{-|z_i|^2/2},\label{}
\end{equation} 
where
\begin{align}
&\phi_{l}(\{z_i\})=\left(\sum_{i_1,i_2,...,i_{N-(N^2-l)}} z_{i_1}z_{i_2}...z_{i_{N-(N^2-l)}}\right)\phi_{N^2-N}(\{z_i\}. 
\nonumber\label{ze}
\end{align}
Each state $\phi_l(\{z_i\})$ has a fixed angular momentum, $l$.  For instance, when $N=3$, $ \Phi_{1,\eta}(z_1,z_2,z_3)=\mathcal{N}_1^{-\frac{1}{2}}\Big(\sum_{l=6}^{9}(-\eta)^{N^2-l}\phi_l(z_1,z_2,z_3)\Big)e^{-(|z_1|^2+|z_2|^2+|z_3|^2)/2}$, and 
\begin{align}
  &\phi_{6}=(z_1-z_2)^2(z_2-z_3)^2(z_3-z_1)^2, \nonumber
\\
&\phi_7 =(z_1+z_2+z_3)(z_1-z_2)^2(z_2-z_3)^2(z_3-z_1)^2,\nonumber\\ 
&\phi_8 =(z_1z_2+z_2z_3+z_3z_1)(z_1-z_2)^2(z_2-z_3)^2(z_3-z_1)^2 , \nonumber\\
&\phi_9=z_1z_2z_3(z_1-z_2)^2(z_2-z_3)^2(z_3-z_1)^2.
\end{align}

It turns out that $\phi_l(\{z_i\})$ corresponds to a Jack polynomial $J^\alpha_{\mathbf{\lambda}
}(\{z_i\})$~\cite{Bernevig2008}. A Jack polynomial~\cite{Lapointe2000}, which has $N$ arguments, $\{z_i\}\equiv\{z_1,z_2,..., z_N\}$, depends on a number of indices. One of them is denoted by $\alpha$. For relevant states in our case,  $\alpha=-2$. For the Moore-Read and other quantum Hall states,  $\alpha$ takes different values. $\lambda$,  a short-hand notation for a set of indices, $\{\lambda_1,\lambda_2, ...,\lambda_N\}$, where $\lambda_1\ge \lambda_2\ge\lambda_3\ge...\ge \lambda_N$, depends on the angular momentum, $l$, and the total particle number, $N$.  If we denote $l$ as $l=N(N-1)+l_r$ with $l_r>0$,
the explicit expression of $\lambda_i$ can be written as
\[ 
\lambda_i = \left\{
\begin{array}{ll}
      2N-2i+1,& i<l_r,  \\
      2N-2i, & i\geq l_r.\\
\end{array} 
\right. 
\]
$\lambda=\{\lambda_1,\lambda_2,...,\lambda_N\}$ specifies one of the monomials in the Jack polynomial, 
\begin{equation}
  m_\lambda=\frac{1}{\prod_{k=0}^{\infty}(n_k!)}\text{Permanant}(z_i^{\lambda_i}),
  \label{<+label+>}
\end{equation}
where $n_k$ is the number of times for the integer $k$ to appear in $\{\lambda_1,\lambda_2, ...,\lambda_N\}$. For instance, $m_{420}=(z_1^4z_2^2z_3^0+z_2^4z_3^2z_1^0+z_3^4z_1^2z_2^0)$. Each of these three integers, $4$, $2$, $0$, shows up in $\lambda=\{4,2,0\}$ once and any other integer does not exist. As such, $\prod_{k=0}^{\infty}(n_k!)=1$ in this example. All other monomials in a Jack polynomial can be derived from $  m_\lambda$ by squeezings. 

\begin{figure} 
  \includegraphics[angle=0,width=0.5\textwidth]{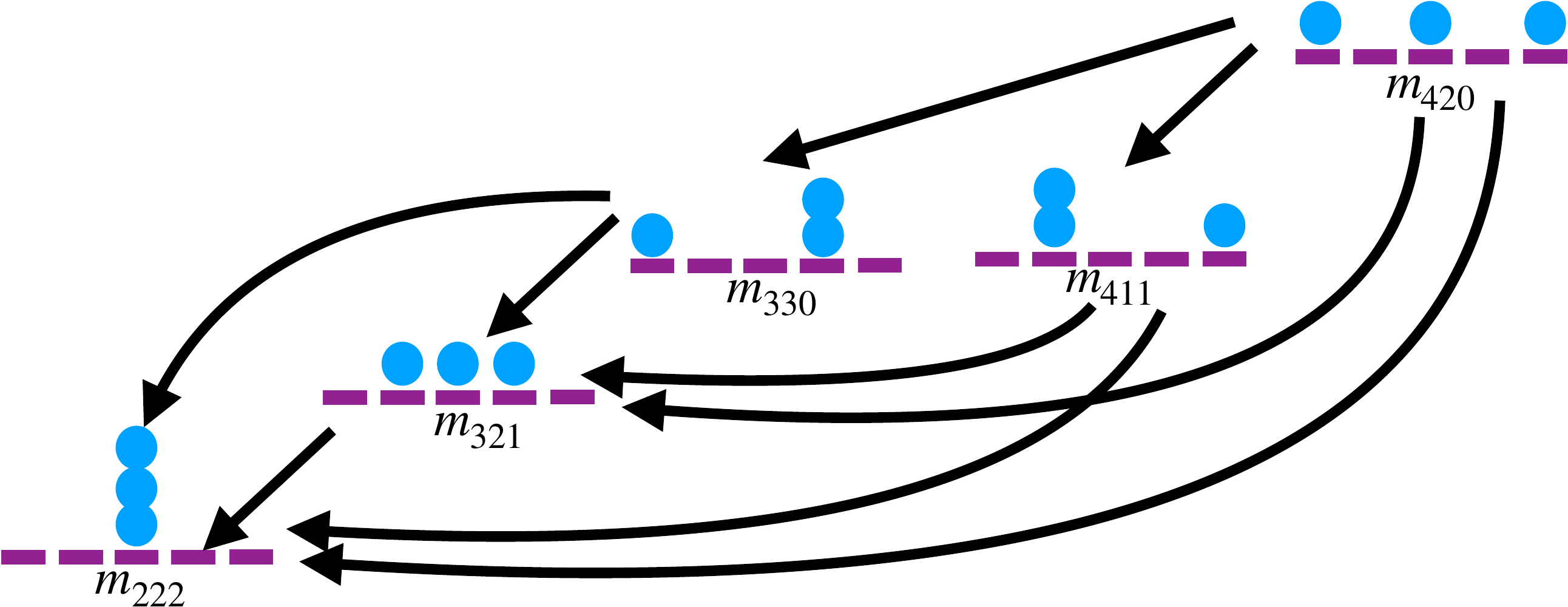}
  \caption{
    Repeatedly applying the squeeze operator to the Jack polynomial of $m_{420}$
    generates all other polynomials for total angular momentum $l=6$.
    } \label{figs1} \end{figure}

The explicit expression of a Jack Polynomial can be written in terms of a determinant,
\begin{equation}\label{Jks}
J_{\lambda}^{\alpha}=\frac{1}{\mathcal{N}_{J_{\lambda}^{\alpha}}} \left|\begin{array}{cccccc}
m_{\lambda^{(1)}} & m_{\lambda^{(2)}} & \cdots & \cdots & m_{\lambda^{(n-1)}} & m_{\lambda^{(n)}} \\
d_{\lambda^{(1)}}-d_{\lambda^{(n)}} & C_{\lambda^{(2)} \lambda^{(1)}} & \cdots & \cdots & C_{\lambda^{(n-1)} \lambda^{(1)}} & C_{\lambda^{(n)} \lambda^{(1)}} \\
0 & d_{\lambda^{(2)}}-d_{\lambda^{(n)}} & & & \cdots & C_{\lambda^{(n)} \lambda^{(2)}} \\
\vdots & 0 & \ddots & & & \vdots \\
\vdots & \vdots & \ddots & \ddots & & \vdots \\
0 & 0 & \cdots & 0 & d_{\lambda^{(n-1)}}-d_{\lambda^{(n)}} & C_{\lambda^{(n)} \lambda^{(n-1)}}
\end{array}\right|,
\end{equation}
where $\lambda^{(n)}\equiv\lambda$.  A squeezing operation, which allows two
particles to exchange a certain amount of angular momentum, changes
$m_{\lambda^{(n)}}$ to another monomial, such as $m_{\lambda^{(n-1)}}$. For
instance, $m_{420}$ becomes $m_{411}$ after applying a squeezing operation in
which the angular momentum of one particle decreases from 2 to 1 and
correspondingly the angular momentum of the other one increases from 0 to 1,
i.e., $\lambda^{(n)}=\{4,2, 0\}$ has become $\lambda^{(n-1)}=\{4,1,1\}$.
Repeating the squeezing operations to $\lambda^{(n)}$, or other states obtained
from previous squeezings, all states in the first row of the above equation are
obtained. Fig.~\ref{figs1}
shows an example of how to obtain all states from
squeezing $m_{420}$.  These $m_{\lambda^{(i)}}$ are ordered based on
$\lambda^{(1)}<\lambda^{(2)}<...<\lambda^{(n)}$.   $\lambda^{(j)}<\lambda^{(k)}$
means that there exist a $i^*$ such that $\lambda^{(j)}_i\le \lambda^{(k)}_i$
for any $i\le i^*$. For instance, the states in Fig.~\ref{figs1} are ordered as $m_{222}, m_{321}, m_{330}, m_{411}, m_{420}$. 

In Eq.(\ref{Jks}), the upper triangle excluding the first row is defined as
\begin{equation}
C_{\lambda^{(j)} \lambda^{(k)}}=\left\{\begin{array}{ll}
\left(\lambda^{(j)}_{i_1}-\lambda^{(j)}_{i_2}\right)\left(\begin{array}{c}
n(\lambda^{(k)}_{i_1}) \\
2
\end{array}\right) & \text { if } \lambda^{(k)}_{i_1}=\lambda^{(k)}_{i_2} \\
\left(\lambda^{(j)}_{i_1}-\lambda^{(j)}_{i_2}\right) n(\lambda^{(k)}_{i_1}) n(\lambda^{(k)}_{i_2})& \text { if } \lambda^{(k)}_{i_1}\neq\lambda^{(k)}_{i_2}
\end{array}\right.
\end{equation}
where  $\lambda^{(k)}_{i_1}=\lambda^{(j)}_{i_1}-k, \lambda^{(k)}_{i_2}=\lambda^{(j)}_{i_2}+k$, 
$0\leq \lambda^{(j)}_{i_2}<\lambda^{(j)}_{i_1}$ and $0<k\le (\lambda^{(j)}_{i_1}-\lambda^{(j)}_{i_2})/2$. As shown in Fig.~\ref{figs1}, $\lambda^{(j)}_{i_1}$ and $\lambda^{(j)}_{i_2}$ are the ones in $\lambda^{(j)}=\{\lambda^{(j)}_{1}, \lambda^{(j)}_{2},...,\lambda^{(j)}_{N}\}$ that change to $\lambda^{(k)}_{i_1}$ and $\lambda^{(k)}_{i_2}$ in a squeezing from $\lambda^{(j)}$ to $\lambda^{(k)}$.  $n(\lambda^{(k)}_{i_1})$ is the number of times that $\lambda^{(k)}_{i_1}$ shows up in $\lambda^{(k)}=\{\lambda^{(k)}_{1}, \lambda^{(k)}_{2},...,\lambda^{(k)}_{N}\}$.

The final next-to-diagonal line is defined as
\begin{equation}
d_{\lambda^{(j)}}-d_{\lambda^{(k)}}=\sum_{i=1}^{N}\left(\frac{\alpha}{2}\left((\lambda^{(j)}_{i})^{2}-(\lambda^{(k)}_{i})^{2}\right)-i\left(\lambda^{(j)}_{i}-\lambda^{(k)}_{i}\right)\right).
\end{equation}
Finally, the normalization factor, $1/\mathcal{N}_{J^\alpha_\lambda}$ is chosen such that 
$J^\alpha_\lambda=m_{\lambda^{(n)}} +
\sum_{j=1}^{n-1}c_j m_{\lambda^{(j)}}$, i.e., the coefficient in front of $m_{\lambda^{(n)}}$ is one.

As an example,  the explicit expression of $J^2_{420}$ is written as, 
\begin{equation}
J^2_{420}=\frac{1}{\mathcal{N}_{J^2_{420}}}
\left|
\begin{array}{ccccc}
 m_{222}&m_{321}&m_{330}&m_{411}&m_{420} \\
 -4 & 6 & 0 & 0 & 12 \\
 0 & -4 & 3 & 3 & 4 \\
 0 & 0 & -1 & 0 & 2 \\
 0 & 0 & 0 & -1 & 2 \\
\end{array}
\right|
  \label{<+label+>}
\end{equation}
Compare the above equation to $\phi_6$ that we previously obtained, we find that $\phi_6=J^2_{420}$. 

Similarly,  we could conclude that  $\phi_7=J_{520}$, $\phi_8=J_{530}$, $\phi_9=J_{531}$.

\section{Mapping initial state to monimial basis}

we start from an initial state
\begin{equation}
  \Psi_{\text I,1}=\mathcal{N}_{\text
  I,1}^{-\frac{1}{2}}\prod_{i=1}^N\left(\sum_{m=1}^{N} c_m
  {z_i^m}\right)\prod_i e^{-|z_i|^2/2}, c_m=1.\label{psiI1}
\end{equation}
and project to different angular momentum channels
\begin{equation}
  \Psi_{\text I,1}=\mathcal{N}_{\text
  I,1}^{-\frac{1}{2}}(\sum_l\psi_l)\prod_i e^{-|z_i|^2/2}.
\end{equation}

We obtain that $\psi_l=\sum_\lambda \prod_i d_{\lambda_i} m_\lambda$.

For example for $N=3$,
\begin{align}
  \psi_6&=d_2^3 m_{222}+d_3 d_2 d_1 m_{321}+d_3^2 d_0 m_{330}\\
  \psi_7&=d_3^2 d_1 m_{331}+d_3 d_2^2 m_{322}\\
  \psi_8&=d_3^2 d_2 m_{332}\\
  \psi_9&=d_3^3 m_{333}.
  \label{<+label+>}
\end{align}
Though the initial state could couple to angular momentum less than 6, we can
safely ignore them since they have zero overlap with the final state.
Finally, combining the equations
\begin{align}
  \braket{\psi_6|J_{\lambda(6)}}&=(-\eta)^3O_c  \braket{J_{\lambda(6)}|J_{\lambda(6)}}\\
  \braket{\psi_7|J_{\lambda(7)}}&=(-\eta)^2O_c  \braket{J_{\lambda(7)}|J_{\lambda(7)}}\\
  \label{l8}
\braket{\psi_8|J_{\lambda(8)}}&=(-\eta)O_c  \braket{J_{\lambda(8)}|J_{\lambda(8)}}\\
  \braket{\psi_9|J_{\lambda(9)}}&=O_c  \braket{J_{\lambda(9)}|J_{\lambda(9)}}
  \label{l9}
\end{align}
and with the condition $d_3=1$, we can solve for $O_d$, and $d_i$.
We solve for $O_d$ first from Eq.~(\ref{l9}), then solve for $d_2$ from
Eq.~(\ref{l8}), etc. The solution is guaranteed because they are linear in each
step. We obtain $O_d=-(9/220),d_0=-((41991
\eta^3)/332750),d_1=((3303 \eta^2)/6050), d_2=-((81 \eta)/55)$. Note, 
$O_d$ differ the overlap $\braket{\Psi_{I,1}|\Phi_{1,\eta}}$ by a normalization factor $(\mathcal{N}_1\mathcal{N}_{1,\eta})^{1/2}$.

\section{Generalization to two quasi-hole state in the main text}
The same analysis could be performed for the two quasihole state. 
We need to replace $\lambda_i$ by $\lambda_i+1$. Once we express $\Phi_{1,\eta}(\{z_i\})$, the state with one quasi-hole, in terms of Jack Polynomials, we could immediately rewrite the state with two quasi-holes, $\Phi_{2,\eta}(\{z_i\})$ also in terms of Jack Polynomials, since $\Phi_{2,\eta}(\{z_i\})\sim(\prod_{i}z_i)\Phi_{2,\eta}(\{z_i\})$.

The normalization constants $N(\lambda)$ changes but everything else, including
the total number of basis and the coefficients in front of the basis remains the same.
Likewise, generalization to cones, e.g., replace $\lambda_i$ by $3\lambda_i$ also only changes the normalization factors. Using Jack polynomial, we are able to calculate the coefficients for up to $N=8$.
The coefficients required to generate the one-quasi-hole and two-quasi-hole
state for $N$ up to 4 is summarized in table~\ref{tab1}.

\begin{table*} \caption{  The coefficients $d_i$ ($c_i$) are for the wave functions with one quasi hole
   exicitation at $\eta$ (one quasi hole excitation at origin and another at
   $\eta$). $O_d$ ($O_c$) is the overlap of the initial wave function and the
   final state.
 }
 \begin{tabular}{ l| C C  C  C  C|  c c  c  c  c}
   $N$ & O_d & d_0 & d_1 &d_2 & d_3    & $O_c$ & $c_1$ & $c_2$ &$c_3$ & $c_4$   \\
   2 & -\frac{2}{7} & \frac{4 \eta^2}{49} & -\frac{8 \eta}{7}  &   &
   &$3/10$ & $-4 \eta^2/25$ & $-9 \eta/10$ & & \\ 
   3 &-\frac{9}{220} & -\frac{41991 \eta^3}{332750} & \frac{3303 \eta^2}{6050} &
   -\frac{81 \eta}{55}    &
   &$12/257$ & $-138478 \eta^3/16974593$  & $2928 \eta^2/66049$  & $-264 \eta/257$\\
   4 &\frac{8}{2377} & \frac{3571758875792 \eta^4}{861946858349307} &
   -\frac{41724784088 \eta^3}{120873209697} & \frac{20331502 \eta^2}{16950387} &
   -\frac{13504 \eta}{7131} &$\frac{125}{28931}$ & $\frac{-88255919558988767 \eta^4}{1401147243843246242}$ & $\frac{1768215130300 \eta^3}{24215326878491}$ & $ \frac{132909825 \eta^2}{837002761}$ & $\frac{-35075 \eta}{28931}$ \\
  \end{tabular}
 \label{tab1}
\end{table*}

\section{solutions to the master equation}

Here, we use dissipative dynamics on a cone as an example. Discussions can be straightforwardly generalized to a flat plane. 
We consider a cavity supporting 
states with angular momenta $3,6,$ and $9$.  
To simplify notations, the decay rate of the $p$-orbital has been taken as a constant in these angular momentum channels. 
As an example, we set $g/\Omega=4.3/1.5$.
We perform a time-dependent calculation using the master equation and find that an initial 
pure 
state of 2 photons decays into
a density matrix that is a mixture of the vacuum, the one-body dark state $d_l^{\dagger}\ket{vac}$ with
angular momentum $l=3,6, $ and $9$, and a two-body quasihole state
$D^\dagger\ket{vac}=\frac{1}{\sqrt{6.2}}d_6^{\dagger}d_6^{\dagger}\ket{vac}-\sqrt{\frac{2.1}{3.1}}d_3^{\dagger}d_9^{\dagger}\ket{vac}$.
Here $\lambda=\Omega/\sqrt{g^2+\Omega^2}=0.329$, and $d_l^{\dagger}=\lambda
a_l^{\dagger}-\frac{g\lambda}{\Omega}r_l^{\dagger}$. 
To be more explicit,  the 
density matrix of the final 
steady state  reads
$c_{vac}\ket{vac}\bra{vac}+c_{3}d_3^\dagger \ket{vac}\bra{vac} d_3+
  c_{9} d_9^\dagger \ket{vac}\bra{vac} d_9+c_{6}d_6^\dagger \ket{vac}\bra{vac} d_6
+c_{tb} D^\dagger \ket{vac}\bra{vac} D$.
Because the initial two photons have angular momentum 6, $c_{3}=c_{9}$.
There are no off diagonal terms in the final density matrix.

\begin{figure} 
  \includegraphics[angle=0,width=0.95\textwidth]{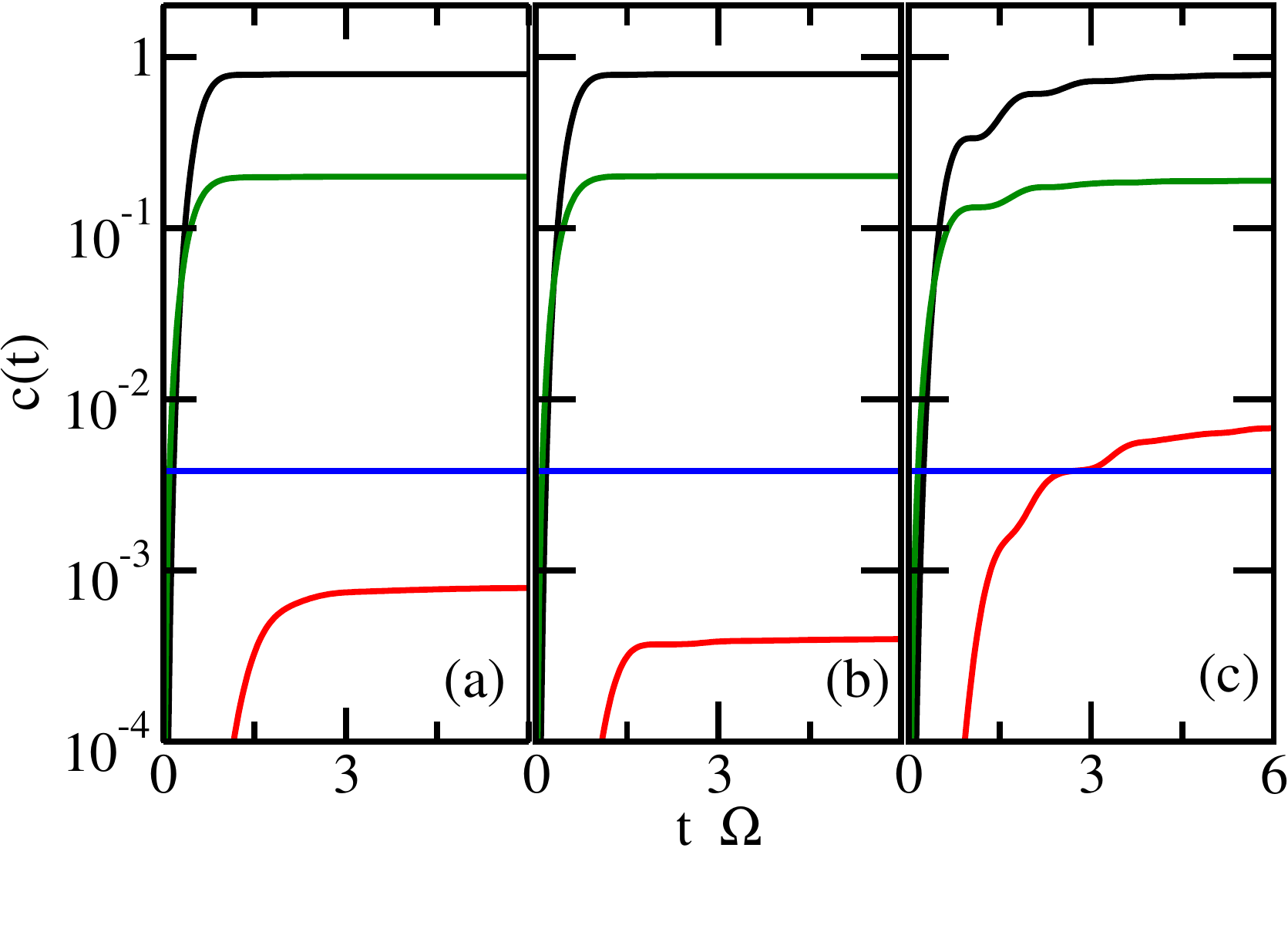}
  \caption{
    The black, green, red, and blue lines
    shows the probability
    $c(t)$ for the diagonal
    density matrix elements $\ket{vac}\bra{vac}$,
    $d_6^\dagger\ket{vac}\bra{vac}d_6$,
    $d_3^\dagger\ket{vac}\bra{vac}d_3$,
    $D^\dagger\ket{vac}\bra{vac}D$ as a function of time.
    $d_9^\dagger\ket{vac}\bra{vac}d_9$ is the same as
    $d_3^\dagger\ket{vac}\bra{vac}d_3$.
    Panel (a) is for $\gamma=2.3\Omega$ and $U_0=74\Omega$.
    Panel (b) is for $\gamma=2.3\Omega$ and $U_0=111\Omega$.
    Panel (c) is  for $\gamma=1.15\Omega$ and $U_0=111\Omega$.
    Changing the interaction energy or decay rate do not change the two-body
    dark state at all but they affects the one-body states and vacuum
    state.
    From (a) to (b), changing $U$ does not affect the decay to the one-body
    sector much, which is mainly governed by the decay rate in the $p$-orbitals.
    } \label{figs2} \end{figure}

For convenience, we define anti-dark state $\tilde{d}_l^\dagger=\frac{g\lambda}{\Omega}a_l^{\dagger}+\lambda
r_l^\dagger$.
We 
rewrite the initial photonic state $(a_6^\dagger)^2\ket{vac}$ 
as a superposition of states 
that are proportional to $
(\tilde{d}_6^{\dagger})^2,
\tilde{d}_6^{\dagger} d_6^{\dagger},
(d_6^{\dagger})^2$.
The first part decays completely.
The second part 
decays 
to a one-body dark state of angular momentum 6 with probability of
$2\lambda^2(1-\lambda^2)=0.19$, which is the majority of $c_6$.
The last part 
decays 
to the two-body quasihole state with the probability of $c_{tb}=\lambda^4/3.1=0.38\%$ and an anti quasihole state.
The anti quaishole states decays into the one-body dark states with $l=3,6,9$.
The coefficients $c_{3},c_{9},c_{6}$ 
depend on the interaction strength and the decay rate.
For example, for $\gamma=2.3\Omega$ and $U_0=74\Omega$ [Fig.~\ref{figs2}(a)],
$c_{3}=0.0008$ and
$c_{6}=0.2001$; for $\gamma=2.3\Omega$ and $U_0=111\Omega$ [Fig.~\ref{figs2}(b)], $c_{3}=0.0004$, and $c_{6}=0.201$; for $\gamma=1.15\Omega$ and
$U_0=111\Omega$ [Fig.~\ref{figs2}(c)], $c_{3}=0.007$, and $c_{6}=0.190$.

The lines in Fig.~\ref{figs2} 
show the time-dependent coefficient, $c(t)$, of the density matrix.
The blue lines 
denote $c_{tb}(t)$. Since it is 
the coefficient for the dark state, 
$c_{tb}(t)$ stays as a constant regardless of the interaction
strength and the decay rate.
The red, green, black lines 
denote 
$c_{3}$, $c_{6}$, and 
$c_{vac}$, respectively.
Changing the interaction strength [Fig.~\ref{figs2}(a)] or the decay rate [Fig.~\ref{figs2}(c)] 
leads to changes in
$c_{3}$, $c_{6}$, and $c_{vac}$.

\end{document}